\documentstyle[emulateapj,epsf]{article}

\submitted{To appear in the Astrophysical Journal.}

\lefthead{Montes et al.}
\righthead{Radio Observations of SN 1979C}

\def\mdot{\dot{M}}
\def\Msun{M_{\sun}}
\def\msun{M_{\sun}}

\begin{document}

\title{Radio Observations of SN~1979C: Evidence for Rapid Presupernova
Evolution}

\author{Marcos J. Montes}
\affil{Naval Research Laboratory, Code 7212, Washington, DC 20375-5320; 
montes@rsd.nrl.navy.mil}
\authoraddr{Code 7212, Washington, DC, 20375-5320}

\author{Kurt W. Weiler}
\affil{Naval Research Laboratory, Code 7214, Washington, DC 20375-5320; 
weiler@rsd.nrl.navy.mil}
\authoraddr{Code 7213, Washington, DC, 20375}

\author{Schuyler D. Van Dyk}
\affil{IPAC/Caltech, Mail Code 100-22, Pasadena, CA 91125; vandyk@ipac.caltech.edu}

\author{Nino Panagia\altaffilmark{1}}
\affil{Space Telescope Science Institute, 3700 San Martin Drive, Baltimore, MD 
21218; panagia@stsci.edu}

\author{Christina K. Lacey}
\affil{Naval Research Laboratory, Code 7214, Washington, DC 20375-5320; 
lacey@rsd.nrl.navy.mil}
\authoraddr{Code 7213, Washington, DC, 20375}

\author{Richard A. Sramek}
\affil{P.O. Box 0, National Radio Astronomy Observatory, Socorro, NM
87801; dsramek@nrao.edu}

\and
\author {Richard Park\altaffilmark{2}}
\affil{Thomas Jefferson High School for Science and Technology,
Alexandria, VA; rpark@mit.edu}

\altaffiltext{1}{Affiliated with the Astrophysics Division, Space
Science Department of ESA.}
\altaffiltext{2}{Now at the Massachusetts Institute of Technology.}

\begin{abstract}
We present new radio observations of the supernova SN~1979C made with
the VLA at 20, 6, 3.6, and 2 cm from 1991 July to 1998 October, which
extend our previously published observations (Weiler et
al.\ \markcite{w86}1986, \markcite{w91}1991), beginning 8 days after
optical maximum in 1979 April and continuing through 1990 December.  We
find that the radio emission from SN~1979C has stopped declining in
flux density in the manner described by Weiler et
al.\ \markcite{w92}(1992), and has apparently entered a new stage of
evolution.  The observed ``flattening,'' or possible brightening, of
the radio light curves for SN~1979C is interpreted as due to the SN
shock wave entering a denser region of material near the progenitor
star and may be indicative of complex structure in the circumstellar
medium established by the stellar wind from the red supergiant (RSG)
progenitor.
\end{abstract}

\section{Introduction}
The study of supernovae (SNe) which are significant sources of radio
emission, known as ``radio supernovae'' (RSNe), provides unique
information on the properties of the progenitor stellar systems and
their immediate circumstellar environments.  In particular, changes in
the density of the presupernova stellar wind established circumstellar
material (CSM) alter the intensity of the radio emission and allow us
to probe the mass-loss history of the supernova's progenitor,
structures in the CSM, and the nature and evolution of the SN
progenitor.

Significant deviation of the radio emission of SNe from standard models
has previously been noted and interpreted as due to a complex CSM
density structure. SN~1979C (Weiler et al.~\markcite{w86}1986,
\markcite{w91}1991) has shown a quasi-periodic variation in its radio
emission which may be due to modulation of the CSM density by a binary
companion (Weiler et al.~\markcite{w92}1992; Schwartz \& Pringle
\markcite{sp96}1996).  SN~1987A, after an initial, faint radio outburst
and rapid decline until 1990 June, is now increasing in radio flux
density due to the SN shock starting to impinge on the inner edges of
the well known, much higher density central ring (Turtle et
al.\ \markcite{t90}1990; Staveley-Smith et al.\ \markcite{s92}1992,
\markcite{s93}1993, \markcite{s95}1995; Ball et
al.\ \markcite{b95}1995; Gaensler et al.\ \markcite{g97}1997).
SN~1980K (Montes et al.\ \markcite{m98}1998) at an age of $\sim 10$
years has experienced a sharp drop in its radio emission far beyond
that expected from models of its previous evolution. More recently,
SN~1988Z has also shown a sharp drop in its radio emission similar to that
of SN~1980K (Lacey et al.\ \markcite{l99}1999).

Recent observations of SN~1979C imply that the shock from SN~1979C
has entered a new, higher density CSM structure different from the
modulated decline previously reported (Weiler et
al.~\markcite{w92}1992).  The radio emission has apparently stopped
declining and has been constant, or perhaps increasing, for the past
eight years.

We have monitored SN~1979C 
[RA$(2000.0)=12^{\rm h}22^{\rm m}58\fs 67\pm 0\fs 01$; 
DEC$(2000.0)= +15^{\circ}47\arcmin 51\farcs 8 \pm0\farcs 2$] 
in NGC 4321 (M 100) since 1979 April 27, and the results through
December 1990 are available in Weiler et al.~(\markcite{w86}1986,
\markcite{w91}1991, \markcite{w92}1992).  Here we present new Very
Large Array\footnote{The VLA is operated by the National Radio
Astronomy Observatory of the Associated Universities, Inc., under a
cooperative agreement with the National Science Foundation.} (VLA)
radio measurements between 1991 July 28 and 1998 October 19.

\vspace{14pt}

\section{Observations}
\begin{figure}
\plotfiddle{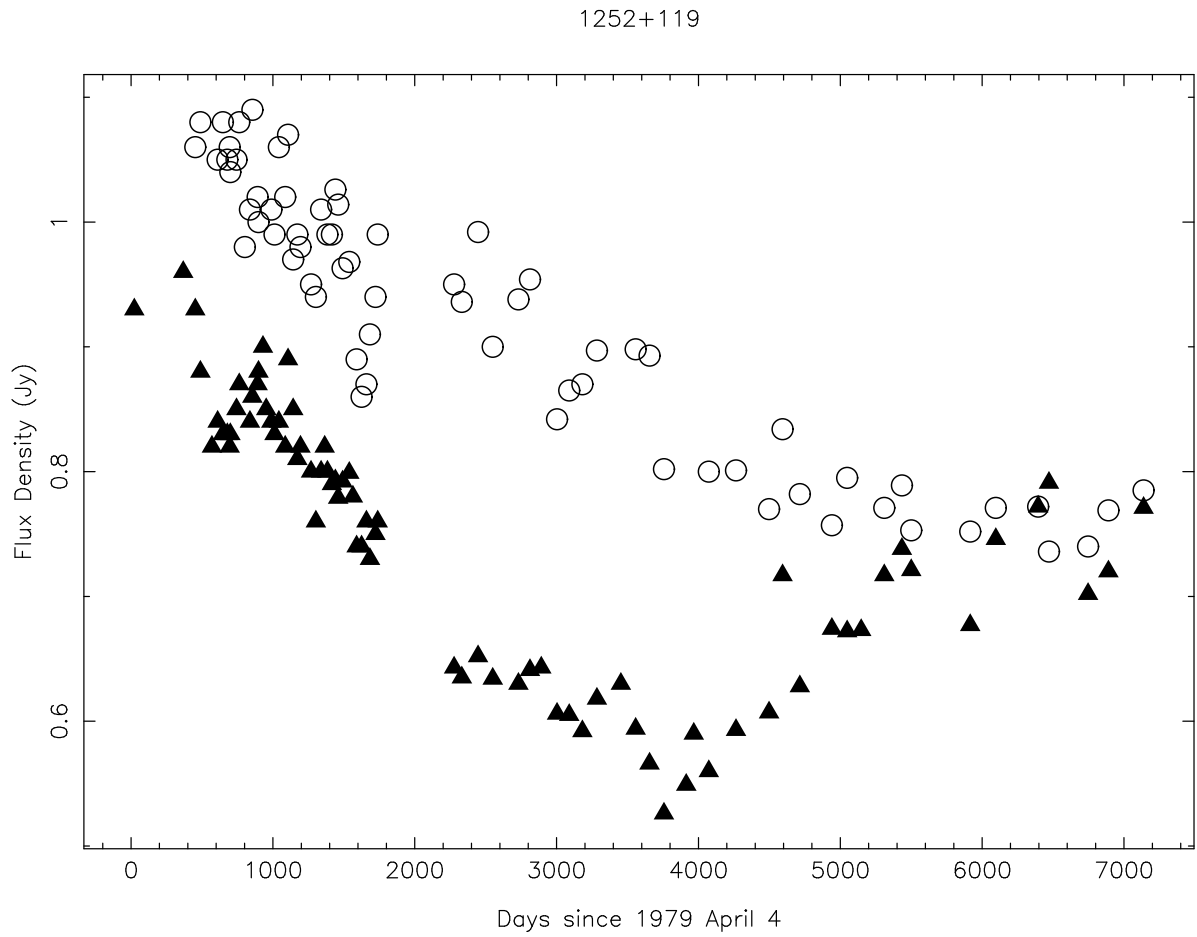}{400pt}{270}{65}{65}{-260}{380}
\caption[1252+119]{Measured flux densities for the secondary calibrator 
1252+119 at 6 cm ({\it filled triangles}) and 20 cm ({\it open
circles}).
\label{fig:cal}} 
\end{figure}

Routine monitoring observations of SN~1979C have been carried out with
the VLA at 20 cm (1.465 GHz) and 6 cm (4.885 GHz) a few times a year
since 1991.  On 1996 December 21 \& 1998 February 10, observations were
additionally made at 3.6 cm (8.435 GHz) and 2 cm (14.965 GHz).  The
techniques of observation, editing, calibration, and error estimation
are described in previous publications on the radio emission from SNe
(see, e.g., Weiler et al.\ \markcite{w86}1986), with particular
reference to SN~1979C in Weiler et al.~(\markcite{w91}1991,
\markcite{w92}1992).  The ``primary'' calibrator remains 3C~286 and is
assumed to be constant in time with flux densities 14.45, 7.42, 5.20,
and 3.45 Jy, at 20, 6, 3.6, and 2 cm, respectively.  The ``secondary''
calibrator, 1252+119, was used as the phase (position) calibrator with
a defined position of RA(1950)=$12^{\rm h}52^{\rm m}07\fs 724$,
DEC(1950)= +$11^{\circ}57\arcmin 20\farcs 82$, and, after calibration
by 3C~286, as the actual flux density reference for the observations of
SN~1979C.  As expected for a ``secondary''
calibrator\footnote{Secondary calibrators are chosen to be compact and
unresolved by the longest VLA baselines. Such compact objects are
usually extragalactic and variable so that their flux density must be
recalibrated from the primary calibrators for each observing session.},
the flux density of 1252+119 has been varying over the years, as can be
seen in both Table \ref{tab:cal} and Figure \ref{fig:cal}.

The flux density measurement error is a combination of the rms map
error, which measures the contribution of small unresolved fluctuations
in the background emission and random map fluctuations due to receiver
noise, and a basic fractional error $\epsilon$ included to account for
the normal inaccuracy of VLA flux density calibration (see, for
example, Weiler et al. \markcite{w86}1986) and possible deviations of
the primary calibrator from an absolute flux density scale.  The final
errors ($\sigma_f$) as listed in Table \ref{tab:data} are taken as
\begin{equation}
\sigma_{f}^{2}\equiv(\epsilon S_0)^2+\sigma_{0}^2 
\label{eq:err}
\end{equation} where
$S_0$ is the measured flux density, $\sigma_0$ is the map rms for each
observation, and $\epsilon =0.05$ for 20, 6, and 3.6 cm, and $\epsilon
= 0.075$ for 2 cm. 

\section{Results}
\begin{figure}
\plotfiddle{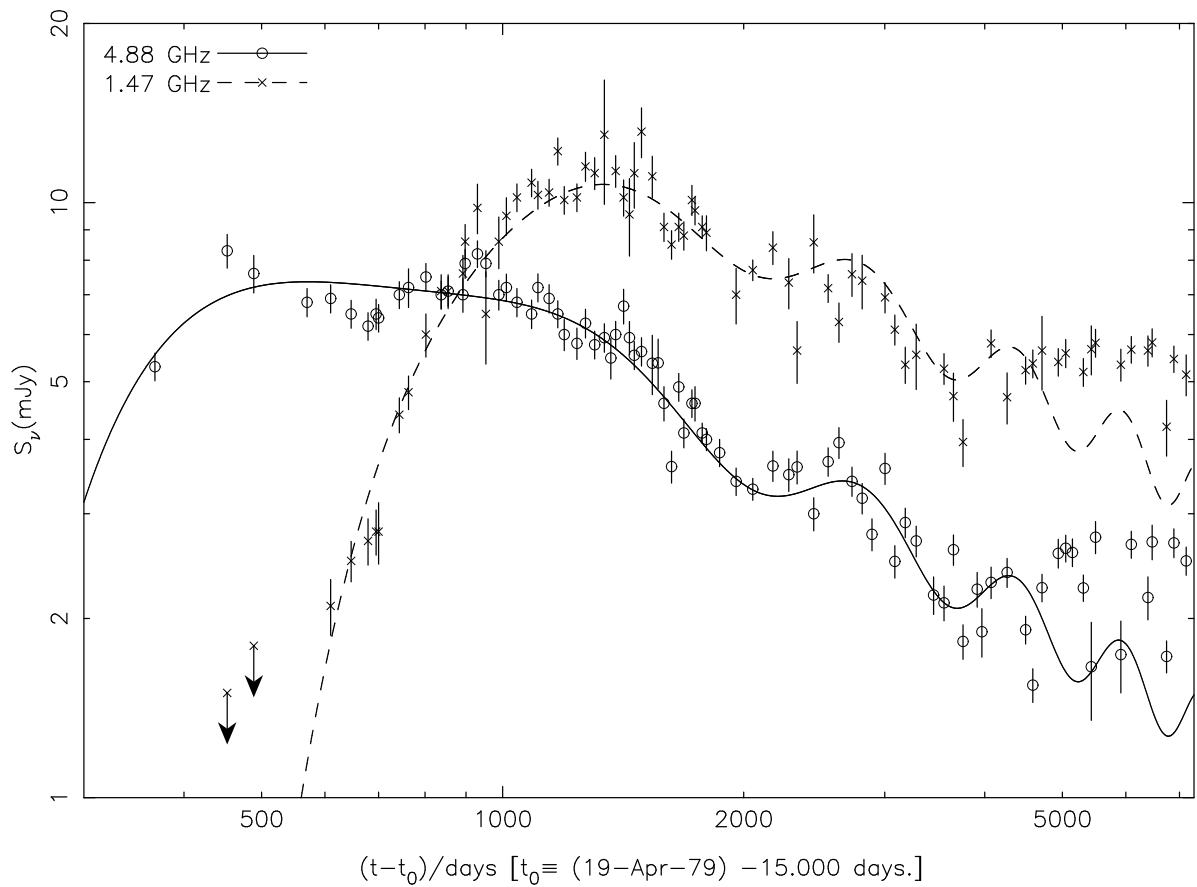}{300pt}{270}{65}{65}{-260}{380}
\caption[SN~1979C light curves]{Radio ``light curves'' for SN~1979C in
NGC 4321 (M100) at wavelengths 20 cm ({\it crosses}) and 6 cm ({\it open
circles}). (Since there are only two new measurements available at 3.6
and 2 cm, they are not plotted here. However, they are shown as
confirming the nonthermal nature of the spectrum in Figure
\ref{fig:spec}.) The data represents about 18 years  of observations
for this object, including the new observations presented in this paper
and previous observations from Weiler et al.~(\markcite{w86}1986,
\markcite{w91}1991).  The curves represent the best-fit model light
curves at 6 cm ({\it solid}), and 20 cm ({\it dashed}), including the
quasi-periodic, or sinusoidal, term proposed by Weiler et
al.\ \markcite{w92}(1992). The best-fit parameters were determined using
only data through 1990 December (day $\sim 4,\!300$).
\label{fig:lc}}
\end{figure}

\begin{figure}
\plotfiddle{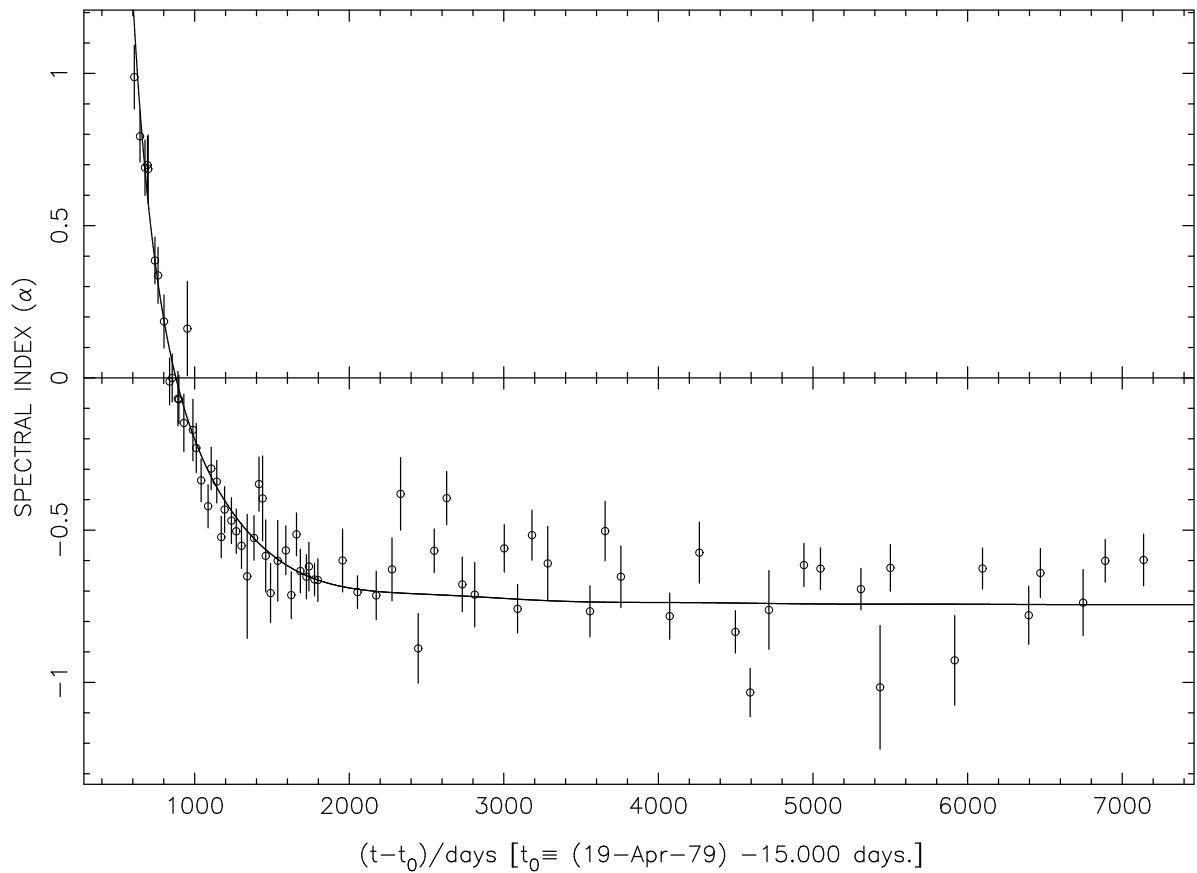}{400pt}{270}{65}{65}{-260}{380}
\caption[$\alpha_{6}^{20}$ for SN~1979C]{Evolution of spectral index 
$\alpha$ (where $S\propto\nu^{+\alpha}$) between 20 and 6 cm
for SN~1979C, plotted as a function of time (in days) since
the estimated explosion date of 1979 April 4 (15 days before optical
maximum). The solid line is calculated from the best-fit theoretical
``light curves'' shown in Fig.\ \ref{fig:lc}.  For reference, see 
Figure 3 in Weiler et al.~\markcite{w86}(1986) and
Figure 2 in Weiler et al.~\markcite{w91}(1991).
\label{fig:si}}
\end{figure}

\begin{figure}
\plotfiddle{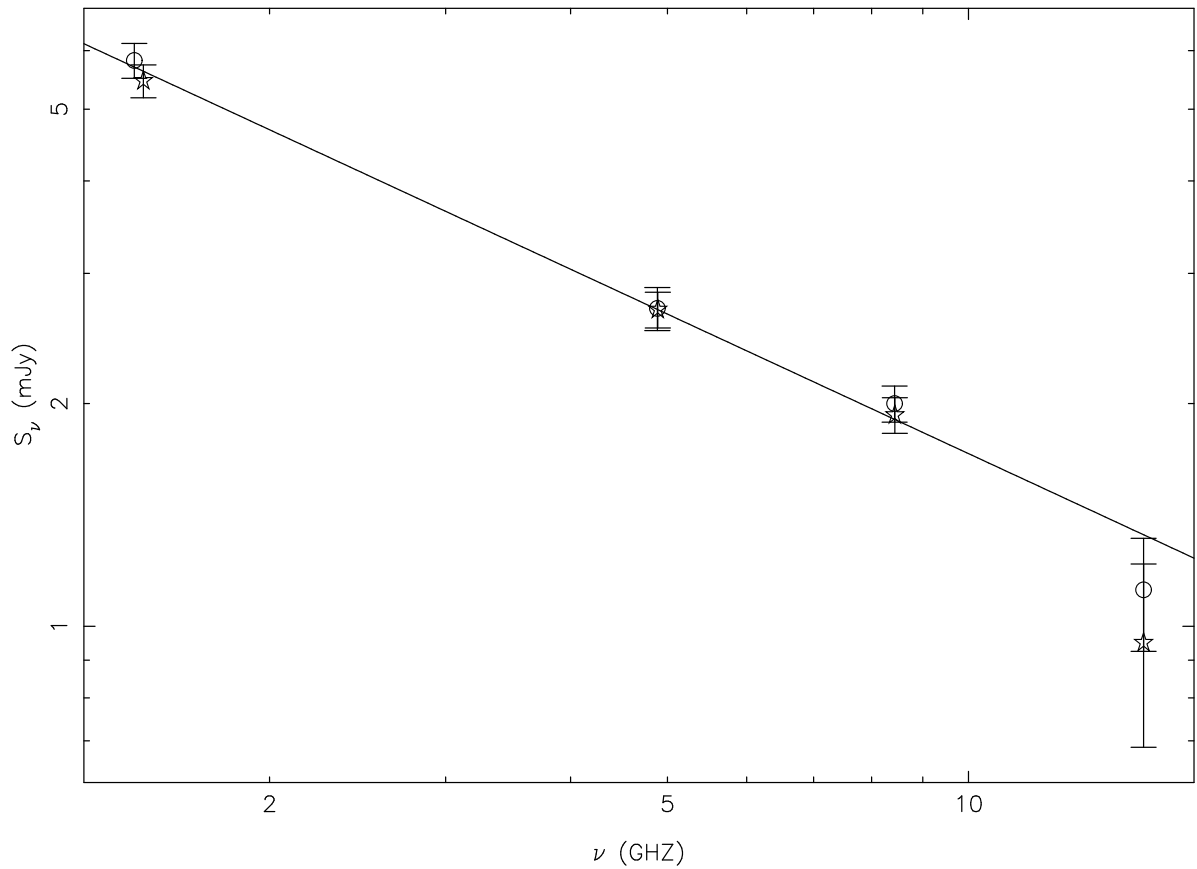}{400pt}{270}{65}{65}{-260}{380}
\caption[1996 Dec. 21 spectrum for SN~1979C]{The radio spectrum for
SN~1979C on 1996 December 21 ({\it open circles}) and 1998 February 10
\& 13 ({\it open stars}) from 20 to 2 cm. Note that, to within the
errors, the spectra are completely optically thin at all frequencies.
The solid line is the best-fit spectrum to the combined data from
both dates. The best-fit spectral index value for the combined data
on these dates is $\alpha=-0.63\pm0.03$.  
\label{fig:spec}}
\end{figure}

Table \ref{tab:data} shows the new flux density measurements from 1991
July 28 through 1998 October 19, while Figure \ref{fig:lc} shows a plot
of all the available 20 and 6 cm data for SN~1979C from 1979 April 27
through 1998 October 19, along with the best-fit model through 1990
December (day $\sim 4,\!300$; Weiler et al.\ \markcite{w92}1992). Figure
\ref{fig:si} shows the evolution of the spectral index between 20 and 6 cm,
$\alpha^{20}_{6}$, for all the available data through 1998 October 19,
and Figure \ref{fig:spec} shows the spectrum at four frequencies for
the 1996 December 21, 1998 February 10, and February 13 observations.

The solid line in Figure \ref{fig:si} is calculated from the best-fit
model to the data through 1990 December and is not
adjusted for the change in evolution since that time.  The solid line
in Figure \ref{fig:spec} is calculated for the best-fit spectrum using
only data from the 1996 December 21, 1998 February 10, and February 13
observations, and may be represented as $S_\nu\propto\nu^{\alpha}$ with
$\alpha=-0.63\pm0.03$.  This spectral index value is clearly flatter
than both the best-fit spectral index after day 4,300, $\alpha_{t>
4,\!300}\sim-0.70$, and the pre-1991 best-fit spectral index of $-0.75$
(see Table 3). Even though for the new measurements since 1990 December
$\alpha_{6}^{20}$ ranges from about $-1 \lesssim \alpha_{6}^{20}
\lesssim -0.6$ (Figure \ref{fig:si}), the spectral index retains a
completely optically thin, nonthermal character, at least over the
wavelength range from 20 to 2 cm (Figure \ref{fig:spec}). It is
apparent that, while the evolution of the radio emission (and
presumably the density structure of the CSM) has changed significantly
since 1990 December, the best-fit spectral index, since becoming
optically thin, has remained fairly constant with $\alpha_{t<4,\!300}\sim
-0.75$ (before day 4,300; Figure \ref{fig:si}) and $\alpha_{t>
4,\!300}\sim-0.70$ (after day 4,300), although short-term deviations such as
that seen in Figure \ref{fig:si} ($\alpha = -0.63$) appear to be real.

The fluctuations seen in Figure \ref{fig:lc} in the 6 cm data (relative
to the fairly constant 20 cm data) after day $\sim 4,\!300$ also appear
to be real. This data was reduced independently by three (and in some
cases four) of the authors, all with similar results. The fluctuations
are not correlated with fluctuations in the secondary calibrator
1252+119 or VLA array configuration. Higher frequency observations
which would assist in the interpretation of this data are,
unfortunately, available on only two dates (1996 December 21 \& 1998
February 13) and have no long-term monitoring values with which to
compare.

\section{Parameterized model}
\label{sec:par}
Previous work on RSNe (see, e.g., Weiler et
al.\ \markcite{w86}1986, \markcite{w91}1991) has shown that the radio
emission can be reasonably well described in its gross properties by a
parameterized model of the form
\begin{equation} 
S {\rm (mJy)} = K_1 {\left({\nu} \over {\rm 5~GHz}\right)^{\alpha}}
{\left({t - t_0} \over {\rm 1~day}\right)^{\beta}}
e^{-{\tau}} \ ,
\label{eq:s}
\end{equation}
where
\begin{equation} 
\tau = K_2 {\left({\nu} \over {\rm 5~GHz}\right)^{-2.1}} {\left({t -
t_0} 
\over {\rm 1~day}\right)^{\delta}} 
\label{eq:tau}
\end{equation}
where $K_1$ and $K_2$ correspond formally to the flux density (in mJy)
and uniform absorption, respectively, at 5~GHz one day after the
explosion date, $t_0$; $\alpha$ is the nonthermal spectral index of the
synchrotron emission; and $\beta$ is the decline rate of the radio
emission after maximum.  The term $e^{-{\tau}}$ describes the
attenuation of a local, external medium that uniformly covers the
emitting source (``uniform external absorption'') and is assumed to be
purely thermal, ionized hydrogen with frequency dependence
$\nu^{-2.1}$.  From the Chevalier
(\markcite{c82a}1982a,\markcite{c82b}b) model for radio emission from
SNe, this CSM is assumed to have radial density dependence
$\rho\propto r^{-2}$ and to have been established by a constant
mass-loss rate, $\mdot$, constant speed wind, $w$, from a red
supergiant progenitor.  The parameter $\delta$ describes the time
dependence of the optical depth for this local, uniform medium, and
$\delta\equiv \alpha - \beta -3$ is specified in the Chevalier model
(Chevalier \markcite{c84}1984). For an undecelerated SN shock,
$\delta=-3$ is appropriate (Chevalier \markcite{c82a}1982a).

Chevalier (\markcite{c82a}1982a) also determined the dependence of the
radio luminosity on the mass-loss rate ($\mdot$) and progenitor wind
velocity ($w$) or, equivalently, to the average CSM density
($\rho_{\rm CSM} \propto\mdot/w$), as
\begin{equation}
 L \propto \left( \frac{\mdot}{w}\right)^{(\gamma-7+12 m)/4}, 
\label{eq:L}
\end{equation}
where $\gamma=-2\alpha+1$ is the power law of the relativistic electron
energy distribution and $m=-\delta/3$ describes the time dependence of
the self-similar evolution of the shock radius, $R\propto t^m$. In this
model, the magnetic energy density and the relativistic energy density
both scale as the total post-shock energy density ($\propto \rho_{\rm CSM}
V_{\rm shock}^{2}$ where $V_{\rm shock}= dR/dt$), and both the magnetic
field amplification efficiency and the particle acceleration efficiency
remain constant as the SN evolves.

Weiler et al.\ \markcite{w92}(1992) noted periodic features in the
radio light curves of SN~1979C. Since the spectral index was not
affected by the observed flux density oscillations, they concluded that
the variations were due to emission efficiency changes caused by
modulations in the CSM density structure, rather than being due to
optical depth effects.  They, therefore, introduced a modification to
Equation \ref{eq:s} by multiplying the emission term $K_1$ by the
sinusoidally-varying modulation
\begin{equation}
\left\{ 1 + A\sin\left[ 2\pi B\left(\frac{t-t_0}{\rm{1\ day}}\right) +
C\right]\right\}^{-(\alpha +2\delta +3)/2} \ .
\label{eq:K1mod}
\end{equation}

Since a radial density modulation of the CSM will also affect the optical
depth $\tau$ ($\tau\propto\rho^{2}_{\rm CSM}\propto(\mdot/w)^2$),
Equation \ref{eq:tau} was modified by multiplying the absorption
term $K_2$ by
\begin{equation}
\left\{ 1 + A\sin\left[ 2\pi B\left(\frac{t-t_0}{\rm{1\ day}}\right) +
C\right]\right\}^2\ .
\label{eq:K2mod}
\end{equation}
In Equations  \ref{eq:K1mod} and \ref{eq:K2mod}, the expression
\begin{equation}
 A\sin\left[ 2\pi B\left(\frac{t-t_0}{\rm{1\ day}}\right) + C\right]
\label{eq:rhomod}
\end{equation}
represents the deviation of $\mdot/w$ from a constant pre-SN wind
mass-loss rate\footnote{While it is true that the optical depth is due
to the {\it integral} along the line of sight to the radio emitting
region, most of the absorption occurs close to the shock front where
the density is the greatest. The fractional errors in $\tau$ from using
our equations (3) and (6) instead of the actual integrated expression
for $\tau$ vary roughly sinusoidally with a period $\sim 1/B$, and a
magnitude $\sim 2A$. At the epoch we are concerned with, $\tau\ll1$,
thus the fractional errors in $e^{-\tau}$ are much smaller than
$2A$.}.
The parameters $A, B,$ and $C$ define the sinusoidal variation of
$\rho_{\rm CSM}$ ($\propto\mdot/w$), where $A$ represents the
fractional amplitude of the density modulation, $B$ represents its
frequency (in cycles day$^{-1}$), and $C$ represents its phase lag (in
radians).

The best-fit model parameters to the pre-1991 ($t-t_0 < 4,\!300$\ days)
data are listed in Table \ref{tab:fit}. This fit is slightly different
from that of Weiler et al.\ \markcite{w92}(1992), due to improved
fitting software, but agrees with that work to within the
uncertainties.  The errors in these new fitting parameters were
estimated using a bootstrap procedure (Press et
al.\ \markcite{p92}1992). Bootstrap procedures use the actual data sets
to generate thousands of synthetic data sets that have the same number
of data points, but some fraction of the data is replaced by duplicated
original points. The fitting parameters are then estimated for these
synthetic data sets using the same algorithms that are used to
determine the parameters from the actual data. The ensemble of
parameter fits is then used to estimate errors for the parameters by
examining number distributions for the parameter in question. The
errors in the fitting parameters in Table \ref{tab:fit} correspond to the
values with 15.85\% and 84.15\% (i.e., $\pm 1\sigma$ for a Gaussian
distribution), respectively, of the cumulative distribution for each parameter. 

\section{Discussion}
\subsection{Interpretation}
As can be seen in Figure \ref{fig:lc}, the new measurements for
SN~1979C since 1991 July 28 do not fit the model which described the
data relatively well through 1990 December.  Since that time, the flux
densities at all frequencies appear to have stopped declining and
perhaps even started increasing.  Accepting our previous interpretation
that a constant spectral index indicates an unchanged emission
mechanism with constant efficiencies, this flattening implies that the
shock wave of the SN is now interacting with a higher density
structure in the CSM.

Under this scenario, the flattening of the radio light curve may imply
either the presence of denser circumstellar material (\S
\ref{sec:den}), or perhaps a geometrical solution, such as the presence
of a flared disc or ring (\S \ref{sec:disc}).
Such dense shells or rings, similar to those we propose, are present
around early type stars, as indicated by observations of luminous blue
variable (LBV) stars (see, e.g., Nota et al.\ \markcite{n95}1995), the
recent detections of rings around B type supergiants (see, e.g., Brandner
et al.\ \markcite{b97}1997), and even by SN~1987A with its prominent
circumstellar ring.

Another example of a SN surrounded by a circumstellar shell
denser than a normal RSG wind is provided by SN~1978K, for which Montes
et al.\ (\markcite{m97}1997) have found evidence of a constant
free-free absorption that produces a characteristic low-frequency
curvature in the typical synchrotron radio spectrum of the SN.
The presence of a discrete shell of ionized gas surrounding SN~1978K
has been confirmed by Chu et al.\ (\markcite{c99}1999) through
spectroscopic measurements showing narrow H$\alpha$ and [N II] emission
lines with widths and intensities characteristic of the environment
around LBV stars.

Additionally, there is evidence from optical observations of SN~1979C
that the H$\alpha$ and [O I] emission lines are also not decreasing,
but have remained flat or perhaps increased slightly (Fesen et
al.\ \markcite{f99}1999).  The optical emission arises from a different
emission mechanism than the radio emission and at a different region of
the SN shock, providing a separate line of evidence that SN~1979C has
encountered a higher density region of the circumstellar medium.

Furthermore, recent results from simulations of young SNRs interacting
with higher density clouds of varying sizes in an otherwise uniform CSM
(Jun \& Jones \markcite{j99}1999 and references therein) have shown
that radio emission increases due to the increased ambient density and
turbulent magnetic field amplification after a SNR/cloud  interaction
begins. The simulations also showed changes in the efficiencies over
time arising from the SNR/cloud interaction, and how, in such
interactions, the efficiencies, magnetic fields, and densities do not
scale in the simple manner described by the Chevalier
(\markcite{c82a}1982a,\markcite{c82b}b) model.

However, it must be noted that a change in the observed radio emission
could be caused by a change in the efficiency of particle acceleration
without significant CSM density change. Since the efficiency of
particle acceleration is not well understood, an efficiency change may
or may not cause a noticeable change in the spectral index.  If we take
the $\eta$ to be the ratio between the number of electrons that are
accelerated to the total number of electrons, then $\eta$ is linearly
proportional to the measured flux density. While we may not know
$\eta$, we do know that the ratio of the measured flux density to that
expected from the Chevalier model. If we assume that {\em only} $\eta$
changes, while the densities and magnetic fields behave as expected
from the Chevalier model, then $\eta_{7100\mbox{
days}}/\eta_{4300\mbox{ days}}= S_{\rm measured}/S_{\rm model} \sim
1.7$. This would require that the acceleration efficiency had remained
roughly constant until $\sim 4,\!300$ days after explosion, then
continually increase, being some $\sim 70\%$ greater by day $\sim
7,\!100$ than at day $\sim 4,\!300$.  Nevertheless, given the  optical
evidence of an increase in the circumstellar density, which does not
rely on particle acceleration, we interpret the deviation of the radio
emission of SN~1979C from the previously assumed model as due to local
density enhancement with the particle acceleration efficiency
relatively unchanged.

\subsection{Nature of the Necessary Density Enhancement}
\subsubsection{CSM Density Increase} \label{sec:den}
A denser CSM may either be an isotropic shell or be
in the form of condensations which have a small combined covering
factor $\phi= \Omega/4\pi<1,$ relative to the expanding shock front
which is overtaking them.  Since the flux density increase at
$t\simeq7,\!100$ days is about a factor of 1.7 greater than that
expected from the previous best-fit model (Weiler et
al.\ \markcite{w91}1991), and since the radio luminosity has a power-law
dependence on the CSM density ($L_{\rm radio}\propto \rho^{1.82}_{\rm
CSM}$, Chevalier \markcite{c82a}1982a), the implied density enhancement is
a factor of $\sim1.34\phi^{-0.55}$  greater than the expected density for a
$\rho \propto r^{-2}$ constant mass-loss rate, constant velocity wind
established CSM.  Moreover, the increase relative to the best-fit model
behavior is gradual with time, indicating a relatively smooth density
distribution, which decreases with radius more slowly than the expected
$r^{-2}$ behavior.  To produce an increase of density by a factor $\sim
1.34$ between 4,300 and 7,100 days a new radial density profile of
approximately $\rho_{\rm CSM}\propto r^{-1.4}$ is required if $\phi=1$,
i.e., for a uniform density distribution.

If the density enhancement is the result of discrete features, whether
a shell, condensations, or a ring, it is likely to fill a relatively
limited fraction of the space around the SN progenitor, and to be
present over a limited interval of radii ($\Delta R/R < 1$). Thus, this
increase of radio emission could be a transient phase that disappears
in a relatively short time.

The argument for a limited duration to the increased radio emission, at
least for a spherical geometry, is based on the plausible initial mass
of the SN's RSG progenitor. The best-fit parameters for the radio
emission at $t<4,\!300$ days (Table 3) imply a pre-SN mass-loss rate of
$\dot{M}\simeq 1.6 \times 10^{-4}~M_\odot~\mbox{yr}^{-1},$ if one
adopts a pre-SN stellar wind velocity of $10~\mbox{km s}^{-1}$, a shock
velocity of 9,250 km s$^{-1}$ (Bartel \markcite{b91}1991), and an
electron temperature ($T_e$) of $30,\!000~\mbox{K}$ for the CSM.  At a
SN shock velocity of $9,250 ~\mbox{km s}^{-1}$, by 4,300 days after the
explosion the front has reached a distance in the pre-SN wind, with
assumed velocity 10 km s$^{-1}$,  where the material was lost from the
star $\sim 11,\!000$ years before the explosion.  At such a mass-loss
rate ($\mdot$) and wind velocity ($w$) the CSM, through which the shock
has already passed, amounts to $\sim 1.7\msun$, a significant amount of
material, even from a massive progenitor star.  The additional matter
present in the next layers, i.e. between 4,300 and 7,100 days,
corresponds to another $\sim2.2\msun$, making a total mass
$\sim3.9~\msun$ of swept-up CSM by 1998 October.  If the flux density
continues to evolve in the present manner, the amount of engulfed
matter increases rapidly: for example, in another 8 years (i.e., by the
year 2007) the additional swept mass would be $\sim 2.9\msun$, giving a
total swept-up mass of $\sim 6.8~\msun$.

Such a swept wind mass is large, even for a red supergiant star, and it
implies a very massive progenitor.  Estimates of the mass of the
envelope ejected by SN~1979C range from $M_{\rm env}\sim1\Msun$
(Chugai \markcite{c85}1985) to $M_{\rm env}\sim6\Msun$ (Branch et
al.\ \markcite{b81}1981; Bartunov \& Blinnikov \markcite{b92}1992;
Blinnikov \& Bartunov \markcite{b93}1993).  In particular, in the
models by Bartunov \& Blinnikov (\markcite{b92}1992) for SN~1979C, the
CSM has a reasonable density profile, matching the observed $B$ light
curves of SN~1979C quite well. Adopting their value for $M_{\rm
env}$, assuming a stellar remnant mass of $M_{\rm rem}\sim
1.4\Msun$, and using our above estimate for the current value of the
swept-up mass, the initial mass of the progenitor must have been
$M_{0}\gtrsim 11.3\Msun$. Assuming the current evolution of the flux
density continues, by 2007 our estimate would rise to
$M_0\gtrsim14.2\msun$.  While large, this is still consistent with the
estimates of Van Dyk  et al.\ (\markcite{vd99}1999). From their HST
imaging of the SN~1979C environment, they find that the stellar ages in
that environment are consistent with the SN progenitor having an
initial mass of $M_0\approx 17$--$18 \msun$.

\subsubsection{Equatorial wind or disc} \label{sec:disc}
Alternatively, the increase in flux density could be an effect of the
geometry of the CSM.  It is conceivable that at small radii the CSM was
distributed in a disk with constant solid angle. At a radius of $\sim
3.4\times 10^{17} ~\mbox{cm}$ ($0.11$ pc, the radius reached by a 9,250
km s$^{-1}$ shock after 4,300 days) the disk thickness increases so
that it covers a larger solid angle, reaching about twice the initial
solid angle by day 7,100 while maintaining a $\rho \propto r^{-2}$
behavior.  Such a flared geometry would ease  the mass requirement if
the original disk subtended  a relatively small angle ($\Omega$) and is
at a large inclination angle, so as to be seen almost edge-on.  In such
a case the mass-loss rate determined by radio light curve fitting (in
which the rate is derived essentially from the free-free absorption of
the CSM) would only be valid within the solid angle subtended by the
disk, rather than over the full $4\pi$ steradians. Then, the mass
engulfed by day 4,300 would be reduced by a factor $\Omega/4\pi$ and
could be as small as $\sim1/10$ of the estimates in the previous
section, without making the requirement on the viewing inclination
angle too severe.  However, even in such a case, the flux density
flattening observed since day 4,300 requires an increase of the
interacting surface, so that the mass engulfed at later times increases
at the same rate as the flux density does relative to the best-fit model.
For example, if we take an initial solid angle of the disk to be
$\Omega=4\pi/3$, the mass swept by day 4,300 would be $\sim 0.57~\msun$
and the mass swept-up by day 7,100 would be an additional $\sim
0.54~\msun$.  More generally, if we denote with $\phi$ the fraction of
solid angle subtended by the disk in the inner region, the mass swept
by day 4,300 would be $1.71\phi~\msun$ and the mass subsequently swept
up by day 7,100 would be $\sim 1.44\phi\msun$, a large, but much
smaller amount than for the spherical shell case.

\subsubsection{Model discrimination}
A possible discriminant between the two basic possibilities, an
increase of CSM density or an increase of CSM coverage, is the
measurement of free-free absorption at low frequencies.  For the case
of flattening flux density due to a less rapidly declining (as $\rho
\propto r^{-1.4}$ rather than $\rho \propto r^{-2}$) CSM density the
emission measure will begin to decrease as EM $\propto r^{-1.8}$, while
for the case of flattening flux density due to  increasing coverage
factor, the emission measure behavior will be either the same as, or
steeper than the canonical stellar wind $\rho \propto r^{-2}$ and could
vary as much as EM $\propto r^{-3}$ or more.  In temporal terms, the
differing density dependence of the EM implies that up to 4,300 days
the emission measure decreased as EM$\propto t^{-3},$ and that for
$t-t_0 > 4,\!300 $ days its behavior changed to EM$\propto t^{-1.8}$.
However, this behavior can only be tested at those frequencies where
the optical depth was of the order of unity at 4,300 days, so that
observations at lower frequencies than the 1.4 GHz will be required to
distinguish between the two scenarios.

To estimate how low an observing frequency is required to see the
difference, the best-fit model gives an optical depth of
\begin{equation}
\tau(\nu) = 1.45 \left(\frac{t}{1,\!000~\mbox{days}}\right)^{-3}
\left(\frac{\nu}{1~\mbox{GHz}}\right)^{-2.1}\ , 
\end{equation} 
so that at 4,300 days the free-free optical depth was of the order of
unity at 148~MHz and $\sim0.3$ at 250~MHz. Even 330~MHz observations at
the VLA are insufficient as they would probe an optical depth of only $\sim
0.2$ at day 4,300. Thus, testing these models may be impractical, since
no high resolution, high sensitivity radio telescopes currently exist
at such low frequencies.

\subsection{The Possibility of a Clumpy CSM}
We have previously postulated that the CSM around SN~1979C is highly
structured (Weiler et al.\ \markcite{w91}1991, \markcite{w92}1992), and
we and others have found evidence for a clumpy CSM in SN 1986J (Weiler
et al.\ \markcite{w90}1990) and SN 1988Z (Van Dyk et
al.\ \markcite{v93}1993; Chugai \& Danziger \markcite{c94}1994).  We
may also speculate that there is  evidence for a dense clumpy medium
surrounding SN~1979C from the fact that the 1.4 GHz flux densities at
$t>4,\!300$  days remain relatively constant and consistently higher
than predictions from the best-fit model, while the 5 GHz flux density
values fluctuate from as low as the best-fit model extrapolation, to as
high as $0.47$ times the 1.4 GHz flux densities, with a range of
spectral indices from about $-1 \lesssim \alpha_{6}^{20} \lesssim -0.6$
(as shown in Figure 3).  These fluctuations appear real in that they
greatly exceed the estimated measurement errors and seem to have a time
scale of $\lesssim 1$ year. Unfortunately, the paucity of observations
in the 4,300--7,100 day interval makes it hard to test this hypothesis
in detail.

If real, this 5 GHz fluctuation and 1.4 GHz stability could indicate a
``cooling'' time of $\sim 0.5$ year for the relativistic electrons
responsible for the higher frequency emission and an appreciably
greater time constant for lower energy electrons.  This could be
evidence for the presence of a relatively small number of dense clumps
interspersed in the general, stellar wind-generated CSM, but requires a
cooling effect which is a very strong function of frequency.
Synchrotron and inverse Compton losses scale only as $\nu^{-1/2}$, and
Coulomb losses as $\nu^{1/2}$ (see, for example, the analysis of
SN~1993J by Fransson \& Bj\"{o}rnsson \markcite{fb98}1998), so that
producing such a large variation in time scales over such a small
frequency range by these mechanisms is very difficult. Additionally, in
order to produce such losses at all, synchrotron cooling would require
magnetic fields $H\sim 1$ G, much greater than the expected ${\cal
O}(1\mbox{ mG})$ fields.  Unfortunately, there are too few measurements
at 8.4 and 14.96 GHz (only two at each frequency) to determine if the
fluctuations are also seen at higher frequencies.

Other evidence for dense clumps comes from the estimate of high
densities ($n_{e}>10^6~\mbox{cm}^{-3}$) derived by Fesen et al.\
(\markcite{f99}1999) from optical and UV spectroscopy of SN~1979C.
However, it is hard to assess the significance of their result, because
their density estimate is based on a comparison of [O II] and [O III]
line intensities measured at two epochs about 4 years apart, with the
unverified assumption that the line fluxes do not change with time.
Additionally, the optical emission and radio emission probably arise
from different physical media, making it difficult to directly compare
the densities derived by Fesen et al.\ (\markcite{f99}1999) with the
radio results.  Chugai \& Danziger (\markcite{c94}1994) discriminated
between equatorial disc and clump  models for the case of SN~1988Z by
using the intermediate velocity component of the emission lines with
FWHM of $\sim 2,\!000$ km s$^{-1}$.  Fesen et al.\ (\markcite{f99}1999)
propose that the ``spiky'' profiles of several lines from SN~1979C are
suggestive of clumpy emission regions, and individual spikes in the
profiles of [O I]$\lambda$6300 and [O II]$\lambda$7325 in their Figure
5 have roughly the same velocity width as the lines used by Chugai \&
Danziger (\markcite{c94}1994) for the case of SN~1988Z.

The possible presence of clumps in the CSM and the reality of high
frequency fluctuations should be tested. More frequent, multi-frequency
radio observations ($\sim$ every three months) at a number of
frequencies $>1.4$ GHz might establish the nature of the apparent radio
flux density fluctuations, and simultaneous optical/UV observations
would permit an unambiguous determination of the gas density in the
clumps. We have already begun more frequent monitoring of SN~1979C
with the VLA including both higher (8.4 \& 14.9 GHz) and lower (330 MHz)
frequencies.

\section{Conclusions}
Analysis of the radio emission from the Type II RSN~1979C at 20 and 6
cm from 1991 July 28 through 1998 October 19, has shown that its radio
emission has unexpectedly stopped decreasing in flux density and has
flattened, or perhaps begun increasing, while maintaining a relatively
constant spectral index.  Such behavior is in conflict with the
best-fit model parameter predictions, based on an assumed $\rho\propto
r^{-2}$ CSM established by a constant mass-loss rate, constant velocity
wind from the pre-SN star.  We interpret this ``flattening'' to
indicate that the SN shock wave has encountered a new region of CSM
which was formed by the SN's progenitor $\sim 10,\!000$--$15,\!000$
years before the SN explosion.

Interpretation of the data implies that this new region could either be
a higher density shell, which should soon be crossed by the fast
moving shock, or a ``flared'' disk-like structure in the CSM, if the
mass-loss were constrained to a narrow solid angle. Additionally, rapid
radio flux density fluctuations at 5 GHz, which are not present at 1.4
GHz, are interpreted as possible evidence for clumps or large scale
density enhancements in the CSM.

Continued monitoring of SN~1979C at multiple radio frequencies -- which
is ongoing-- is needed to determine the form and duration of this new
phase in the evolution of the radio light curves and, correspondingly,
the structure of this new component of the CSM.  With the longest,
relatively complete, multi-frequency  data set available for the
emission from any supernova, SN~1979C serves as a unique laboratory for
understanding of the evolution of Type II supernova progenitors, their
pre-SN mass-loss history,  and their interactions with their
local environments.


\acknowledgements
We wish to thank the referee, Stephen Reynolds, for his helpful
comments and suggestions. KWW, CKL, \& MJM wish to thank the Office of
Naval Research (ONR) for the 6.1 funding supporting this research.  CKL
additionally thanks the NRC for funding supporting this research.
Richard Park was a participant in the Science and Engineering
Apprenticeship (SEAP) program at the NRL and continued this work as a
senior thesis project at Thomas Jefferson High School of Science and
Technology.  Additional information and data on RSNe can be found on
{\tt http://rsd-www.nrl.navy.mil/7214/weiler/} and linked pages.


\begin{deluxetable}{lcrrrr}
\tablecaption{Measured flux density values
for the secondary calibrator 1252+119\label{tab:cal}}
\tablewidth{0pt}
\tablehead{ & \colhead{Time Since}& & & & \\
\colhead{Observation} &\colhead{SN 1979C}     & \colhead{$S_{20}$} & \colhead{$S_6$} & 
\colhead{$S_{3.6}$} & \colhead{$S_2$} \\
\colhead{Date} & \colhead{Optical Max. (days)}&\colhead{(Jy)} & \colhead{(Jy)} & \colhead{(Jy)}& \colhead{(Jy)} \\
}
\startdata
1979 Apr 19 & $\equiv 0$& \nodata & \nodata & \nodata & \nodata \\
1991 Jul 28 & 4483 & 0.770 & 0.607 &  \nodata &  \nodata \\  
1991 Oct 31 & 4578 & 0.834 & 0.717 &  \nodata &  \nodata \\
1992 Mar 01 & 4700 & 0.776 & 0.628 &  \nodata &  \nodata \\
1992 Oct 13 & 4926 & 0.757 & 0.674 &  \nodata &  \nodata \\  
1993 Jan 28 & 5033 & 0.795 & 0.672 &  \nodata &  \nodata \\
1993 May 07 & 5132 & \nodata& 0.673&  \nodata &  \nodata \\
1993 Oct 17 & 5295 & 0.771 & 0.717 &  \nodata &  \nodata \\
1994 Feb 18 & 5419 & 0.789 & 0.738 &  \nodata &  \nodata \\
1994 Apr 25 & 5485 & 0.753 & 0.721 &  \nodata &  \nodata \\
1995 Jun 15 & 5901 & 0.752 & 0.677 &  \nodata &  \nodata \\
1995 Dec 12 & 6081 & 0.771 & 0.746 &  \nodata &  \nodata \\
1996 Oct 06 & 6380 & 0.772 & 0.772 &  \nodata &  \nodata \\
1996 Dec 21 & 6456 & 0.736 & 0.791 & 0.753    & 0.704 \\
1997 Sep 23 & 6732 & 0.740 & 0.702 &  \nodata &  \nodata \\
1998 Feb 10 & 6872 & \nodata &  \nodata & 0.845 & 0.876 \\
1998 Feb 13 & 6875 & 0.769 & 0.720 &  \nodata &  \nodata \\
1998 Oct 19 & 7123 & 0.785 & 0.771 &  \nodata &  \nodata \\
\enddata
\end{deluxetable}

\begin{deluxetable}{lccrrrrrrrr}
\small 
\tablewidth{0pt}
\tablecaption{New Flux Density Measurements for
SN~1979C\tablenotemark{a}\label{tab:data}}  
\tablehead{
\colhead{} & \colhead{Time} & & & & & & & & &  \\
\colhead{} & \colhead{Since} & \colhead{}   & \multicolumn{8}{c}{Flux Density (mJy)} \\
\colhead{} & \colhead{Optical} & \colhead{} & \multicolumn{8}{c}{\hrulefill} \\
\colhead{Obs.} & \colhead{Max.\tablenotemark{b}} & \colhead{VLA} &
\colhead{$S_{20}$} & \colhead{$\sigma_{20}$} & \colhead{$S_{6}$} &
\colhead{$\sigma_{6}$} & \colhead{$S_{3.6}$} & \colhead{$\sigma_{3.6}$} &
\colhead{$S_{2}$} & \colhead{$\sigma_{2}$} \\
\colhead{Date} & \colhead{(days)} & \colhead{Conf.} & & & & & & & & \\
}
\startdata
\small 
1979 Apr 19 & $\equiv 0$ & \nodata & \nodata & \nodata & \nodata & \nodata & \nodata & \nodata & \nodata & \nodata \\
1991 Jul 28 & 4483       & A       & 5.229   & 0.276   & 1.915   & 0.103   & \nodata & \nodata & \nodata & \nodata \\
1991 Oct 31 & 4578       & B       & 5.365   & 0.289   & 1.546   & 0.101   & \nodata & \nodata & \nodata & \nodata \\
1992 Mar 01 & 4700       & C       & 5.640   & 0.801   & 2.253   & 0.120   & \nodata & \nodata & \nodata & \nodata \\
1992 Oct 13 & 4926       & A       & 5.396   & 0.287   & 2.574   & 0.143   & \nodata & \nodata & \nodata & \nodata \\
1993 Jan 28 & 5033       & A       & 5.582   & 0.304   & 2.625   & 0.136   & \nodata & \nodata & \nodata & \nodata \\
1993 May 07 & 5132       & B       & \nodata & \nodata & 2.583   & 0.138   & \nodata & \nodata & \nodata & \nodata \\
1993 Oct 17 & 5295       & C/D     & 5.190   & 0.281   & 2.251   & 0.115   & \nodata & \nodata & \nodata & \nodata \\
1994 Feb 18 & 5419       & D/A     & 5.670   & 0.535   & 1.660   & 0.311   & \nodata & \nodata & \nodata & \nodata \\
1994 Apr 25 & 5485       & A       & 5.810   & 0.314   & 2.740   & 0.170   & \nodata & \nodata & \nodata & \nodata \\
1995 Jun 15 & 5901       & A       & 5.340   & 0.328   & 1.740   & 0.240   & \nodata & \nodata & \nodata & \nodata \\
1995 Dec 12 & 6081       & B       & 5.663   & 0.074   & 2.664   & 0.139   & \nodata & \nodata & \nodata & \nodata \\
1996 Oct 06 & 6380       & D/A     & 5.640   & 0.335   & 2.170   & 0.170   & \nodata & \nodata & \nodata & \nodata \\
1996 Dec 21 & 6456       & A       & 5.820   & 0.315   & 2.690   & 0.180   & 2.000   & 0.112   & 1.120   & 0.195   \\
1997 Sep 23 & 6732       & CnD     & 4.200   & 0.452   & 1.727   & 0.107   & \nodata & \nodata & \nodata & \nodata \\
1998 Feb 10 & 6872       & D       & \nodata & \nodata & \nodata & \nodata & 1.930   & 0.107   & 0.950   & 0.264   \\
1998 Feb 13 & 6875       & DnA     & 5.460   & 0.279   & 2.680   & 0.149   & \nodata & \nodata & \nodata & \nodata \\
1998 Oct 19 & 7123       & B       & 5.140   & 0.406   & 2.500   & 0.136   & \nodata & \nodata & \nodata & \nodata \\
\enddata
\tablenotetext{a}{For previous measurements, cf.~Weiler et al.~(1986, 1991).}
\tablenotetext{b}{The date of the explosion is taken to be 1979 April 4, 15
days before optical maximum (cf.\ Weiler et al.\ 1986). }
\end{deluxetable}

\clearpage
\begin{deluxetable}{lcc}
\tablewidth{0pt}
\tablecaption{Fitting Parameters for
SN~1979C\tablenotemark{a}\label{tab:fit}} 
\tablehead{
\colhead{Parameter} & \colhead{Value} & \colhead{Deviation
Range\tablenotemark{b}}
}
\startdata
$K_1$(mJy)    &  1710              & 1440--2060\nl
$\alpha$      & $-0.75$            &$-(0.76$--$0.63)$\nl
$\beta$       & $-0.80$            &$-(0.83$--$0.78)$\nl
$K_2$         & $3.38\times 10^7$  & (2.97--3.82)$\times 10^7$\nl
$\delta\equiv\alpha-\beta-3$& $-2.94$&$-(2.96$--$2.92)$\nl
$A$           &$7.3\times 10^{-2}$ & (6.7--8.1)$\times 10^{-2}$\nl
$B^{-1}$(days)& 1570               & 1520--1610 \nl
$C$           &$0.90\pi$           & (0.83--0.99)$\pi$ \nl
$t_0$   & $\equiv$1979 Apr. 4&  \nodata \nl
$\chi^2/$DOF & 2.47          & \nodata  \nl
\enddata
\tablenotetext{a}{Only data through 1990 December are used in the
fitting procedure.}
\tablenotetext{b}{The error estimates for the parameter values are
determined using the bootstrap method, which is described in
\S\ref{sec:par}.}
\end{deluxetable}


\clearpage

\begin{references}
\reference{b95} Ball, L., Campbell-Wilson, D., Crawford, D. F., \& Turtle,
A. J. 1995, \apj, 453, 864 
\reference{b91} Bartel, N. 1991, in Supernovae: The Tenth Santa Cruz Summer
Workshop in Astronomy and Astrophysics, ed. S. E. Woosley (New York:
Springer-Verlag), p. 503
\reference{b92} Bartunov, O. S., \& Blinnikov, S. I. 1992, Pis'ma \azh, 18,
104 (SvA Lett, 18, 43)
\reference{b93} Blinnikov, S. I., \& Bartunov, O. S. 1993, \aap, 273, 106
\reference{b81} Branch, D., Falk, S. W., McCall, M. L., Rybski, P., Uomoto,
A. K., \& Wills, B. J. 1981, \apj, 244, 780
\reference{b97} Brandner, W., Chu, Y.-H., Eisenhauer, F., Grebel, E. K.,
Points, S. D. 1997, \apj, 489, L153
\reference{c82a} Chevalier, R. A. 1982a, \apj, 259, 302
\reference{c82b} ---------. 1982b, \apj, 259, L85
\reference{c84} ---------. 1984, \apj, 285, L63
\reference{c99} Chu, Y.-H., Caulet, A., Montes, M. J., Panagia, N., Van
Dyk, S. D., \& Weiler, K. W. 1999, \apj, 512, L51
\reference{c85} Chugai, N. N. 1985, Pis'ma \azh, 11, 357 (SvA Lett, 11, 148)
\reference{c94} Chugai, N. N., \& Danziger, I. J. 1994, \mnras, 268, 173
\reference{f99} Fesen, R. A., Gerardy, C. L., Filippenko, A. V., Matheson,
T., Chevalier, R. A., Kirshner, R. P., Schmidt, B. P., Challis, P.,
Fransson, C., Leibundgut, B., \& Van Dyk, S. D. 1999, \apj, 117, 125
\reference{cb98} Fransson, C., \& Bj\"{o}rnsson, C.-I. 1998, \apj, 509, 861
\reference{g97} Gaensler, B. M.,  Manchester, R. N., Staveley-Smith, L.,
Tzioumis, A. K.,  Reynolds, J. E., \& Kesteven, M. J. 1997, \apj, 479, 845
\reference{j99} Jun, B.-I. \& Jones, T. W. 1999, \apj, 511, 774
\reference{l99} Lacey, C. K., Weiler, K. W., Van Dyk, S. D., \& Sramek,
R. A. 1999, \baas, 194, \#86.05
\reference{m97} Montes, M. J., Weiler, K. W., \& Panagia N. 1997, \apj, 488,
792
\reference{m98} Montes, M. J., Van Dyk, S. D., Weiler, K. W., Sramek, R. A., \&
Panagia, N. 1998, \apj, 506, 874
\reference{n95} Nota, A., Livio, M., Clampin, M., Schulte-Ladbeck, R. 1995,
\apj, 448, 788
\reference{p92} Press, W. H., Teukolsky, S. A., Vetterling, W. T., Flannery,
B. P. 1992, Numerical Recipes in Fortran (Cambridge:Cambridge
University Press), p. 686
\reference{sp96} Schwartz, D. H. \& Pringle, J. E. 1996, \mnras, 282, 1018
\reference{s92} Staveley-Smith, L., Manchester, R. N.,  Kesteven, M. J.,
Campbell-Wilson, D., Crawford, D. F., Turtle, A. J.,  Reynolds, J. E.,
Tzioumis, A. K., Killeen, N. E. B. K., \& Jauncey, D. L. 1992, \nat, 355, 147 
\reference{s93} Staveley-Smith, L., Briggs, D. S., Rowe, A. C. R., Manchester,
 R. N., Reynolds, J. E., Tzioumis, A. K., \& Kesteven, M. J. 1993, \nat, 366,
166
\reference{s95} Staveley-Smith, L., Manchester, R. N., Tzioumis, A. K.,
Reynolds, J. E., \& Briggs, D. S. 1995, in IAU Colloquium 145: Supernovae and
Supernova Remnants, ed. R. M. McCray (Cambridge:Cambridge Univ. Press),
p. 309 
\reference{t90} Turtle, A. J., Campbell-Wilson, D., Manchester, R. N.,
Staveley-Smith, L., \& Kesteven, M. J. 1990, \iaucirc, 5086
\reference{vd99} Van Dyk, S. D., Peng, C. Y., Barth, J. A., \& Filippenko,
A. F. 1999, \pasp, 111, 313
\reference{v93} Van Dyk, S. D., Weiler, K. W., Sramek, R. A., \& Panagia,
N. 1993, \apj, 419, L69
\reference{w92} Weiler, K. W., Van Dyk, S. D., Pringle, J., \& Panagia,
N. 1992, \apj, 399, 672
\reference{w91} Weiler, K. W., Van Dyk, S. D., Panagia, N., Sramek, R., \&
Discenna, J.  1991, \apj, 380, 161
\reference{w90} Weiler, K. W., Panagia, N., \& Sramek, R. A. 1990, \apj,
364, 611
\reference{w86} Weiler, K. W., Sramek, R. A., Panagia, N., van der Hulst,
J. M., \& Salvati, M. 1986, \apj, 301, 790
\end{references}
\end{document}